\documentclass[twocolumn]{aastex63}
\usepackage[T1]{fontenc}

\usepackage{amsmath}

\graphicspath{{./}{figures/}}

\begin{document}

\title{
Flare-Driven Plasma Dynamics and Elemental Abundance Redistribution
%Element-specific abundance evolution with flare temperature
%Flare-Driven Evaporation and the Redistribution of Elemental Abundances
%Does Flaring Energy Deposition Alter Elemental Abundance Redistribution?
}

\correspondingauthor{Biswajit Mondal}
\email{biswajit70mondal94@gmail.com, biswajit.mondal@nasa.gov}
\author[0000-0002-7020-2826]{Biswajit Mondal}
\affiliation{NASA Postdoctoral Program, NASA Marshall Space Flight Center, ST13, Huntsville, AL, USA}
\author[0000-0002-5608-531X]{Amy R. Winebarger}
\affiliation{NASA Marshall Space Flight Center, Huntsville, AL, 35812, USA}

\begin{abstract}
Since the advent of X-ray and EUV spectroscopy, the discovery of the First Ionization Potential (FIP) effect—where coronal elemental compositions diverge from their photospheric values based on the element's FIP—has remained a key puzzle in solar and stellar astrophysics. These deviations exhibit significant fluctuations during flares, yet their connection to plasma dynamics has remained unclear. Here, we report a clear correlation between temperature-sensitive flaring 
plasma emission and element-specific abundance changes for a solar flare. 
These findings indicate that energy deposition in the chromosphere drives plasma evaporation from different chromospheric heights, modulating elemental abundances. Hydrodynamic simulations support these observations, showing that varying energy deposition magnitudes generate plasma upflows from different chromospheric heights, leading to element-specific FIP fractionation.
%via the ponderomotive force. 
These results provide new insights into the dynamic coupling of flare energy, plasma flows, and abundance variability, with implications for understanding coupling between different atmospheric layers.
\end{abstract}


\section{Introduction}

Understanding the star's elemental composition is fundamental to unraveling how energy is transported from its deep interior, through the outer atmosphere. While the Sun's outer atmosphere, or corona, draw its mass from the lower atmospheric layers, the elemental composition of the corona differs markedly from that of the solar surface, or photosphere. 
In the magnetically closed solar active regions (AR), elements with a low First Ionization Potential (FIP $<$ 10 eV), are selectively enhanced (fractionated) by factors of 3–4, while high-FIP elements retain their photospheric abundances~\citep{Feldman_1992PhyS...46..202F, 2013ApJ...778...69B, Zanna_2022ApJ...934..159D, Mondal_2023ApJ...955..146M}. This compositional variation is not uniform; it changes across atmospheric structures~\citep{Feldman_1993ApJ...414..381F,Widing_1997ApJ...480..400W,Brooks_2017NatCo...8..183B,Doschek_2019ApJ...884..158D,Vadawale_2021ApJ...912L..12V}. This phenomenon, known as the FIP effect, serves as a vital diagnostic tool in solar and stellar astrophysics, providing insights into energy transport and plasma processes.
Highly magnetized stars often exhibit the inverse-FIP effect, where low-FIP elements are depleted in the corona \citep{Testa_2010SSRv..157...37T}. The ponderomotive force theory~\citep{Laming_2004ApJ...614.1063L,Laming_2012ApJ...744..115L} provides a widely accepted explanation for both the FIP and inverse-FIP effects. According to this theory, magnetic waves reflected in the lower atmosphere separate ions from neutrals, driving the observed fractionation in the plasma.

Earlier observations revealed that the FIP effect deviates from its typical behavior near the peak X-ray emission of solar flares~\citep{Sylwester_1984Natur.310..665S,Warren_2014ApJ...786L...2W,Dennis_2015ApJ...803...67D,syama_2020SoPh..295..175N}. \cite{Mondal_2021} further established a clear pattern of abundance variation during multiple small B-class solar flares, showing that in the impulsive phase the abundances of low-FIP elements unexpectedly decline toward photospheric levels before recovering their coronal values in the decay phase. They used the Solar X-ray Monitor (XSM:~\citealp{Mithun_2020SoPh..295..139M, Mithun_2021ExA....51...33M}) onboard the Chandrayaan-2 mission, which provides superior energy resolution, cadence, and sensitivity compared to previous instruments. Similarly, \cite{Mithun_2022ApJ...939..112M,Nama_2023SoPh..298...55N,yamini_2023ApJ...958..190R} found similar abundance variations in C-class and A-class solar flares. These variations have also been confirmed by recently available solar instruments~\citep{Suarez_2023,Ng_2024ApJ...972..123N}.
These transient abundance changes are also observed in stellar flares~\citep{Nordon_2007A&A...464..309N,Karmakar_2023MNRAS.518..900K}, suggesting they are universal in magnetically active stars. However, the underlying mechanisms linking flare-induced plasma dynamics to abundance variations remain unclear.  Few previous studies have suggested flare-driven chromospheric evaporation as a possible cause (e.g., \citealp{Laming_2009ApJ...707L..60L, Mondal_2021}), but direct observational evidence has been lacking.

In this study, we aim to address the gaps in understanding the relationship between flaring plasma dynamics and elemental abundance variations. Our analysis focuses on an M-class solar flare, utilizing X-ray observations from the Solar X-ray Monitor (XSM) and EUV observations from the Atmospheric Imaging Assembly (AIA; \citealp{Lemen_2012SoPh..275...17L}) aboard the Solar Dynamics Observatory (SDO).
Time-resolved spectral analysis of XSM data is used to track plasma temperature, and the abundances of low-FIP elements (Si, Fe) and a mid-FIP element (S). To investigate the connection between abundance evolution and plasma dynamics, we combine the XSM spectral results with high-resolution EUV images from AIA. This integrated approach allows us to directly correlate abundance changes observed by XSM with plasma dynamics captured by AIA.
In addition, we perform a set of hydrodynamic simulations to further explore the coupling between plasma dynamics and abundance variations.

The rest of the paper is organized as follows. Section~\ref{sec-method} describes the methodology, including details of the flare observations (Section~\ref{sec-obs}), time-resolved spectroscopy (Section~\ref{sec-sectroscopy}), and plasma dynamics (Section~\ref{sec-plasmaflows}). The primary results are presented in Section~\ref{sec-results} and discussed in Section~\ref{sec-discussion}. A brief summary is provided in Section~\ref{sec_summary}.

\section{Methods}\label{sec-method}

\subsection{Flare Observations}\label{sec-obs}

The M-class flare, observed on September 17, 2022, at the west limb of the Sun, was detected by both AIA and XSM. 
The XSM captures disk-integrated solar spectra in the 1 -- 15 keV energy range with a one-second cadence and an energy resolution better than 180 eV at 5.9 keV~\citep{Mithun_2021ExA....51...33M}.
Owing to its high sensitivity, it has been effectively used to study coronal X-ray  bright points (XBPs; \citealt{Vadawale_2021ApJ...912L..12V,mondal_2023b}) and ARs~\citep{Mondal_2023ApJ...955..146M,Mondal_2025ApJ...980...75M}. During the flare, however, the XSM spectra were dominated by intense flaring emission~\citep{Mondal_2021}, with only minimal contributions from the rest of the solar disk. 
Meanwhile, AIA captures image sequences in multiple extreme-ultraviolet (EUV) passbands sensitive to a range of temperatures with a spatial resolution of $\sim$1.5" and a temporal cadence of 12 s.  AIA Level 1 data were deconvolved with the instrument's point spread function (PSF) and processed to Level 1.5 using the standard procedures in SunPy~\citep{sunpy_software_stuart_j_mumford_2024_13948147}.

Several of the AIA passband have a bi-modal temperature response, meaning they are sensitive to both hot and cool plasma.  In particular, the AIA 94\,\AA\ passband encompasses dominant emission lines from both Fe VIII and Fe XVIII.  In this paper, we utilized the empirical  method developed by \cite{2013A&A...558A..73D}, which removes cooler line contamination by combining appropriately scaled AIA 171~{\AA} and 211~{\AA} data.  The remaining emission is expected to be dominated by Fe XVIII. 

\begin{figure}[!h]
\begin{center}
    \includegraphics[width=0.48\textwidth]{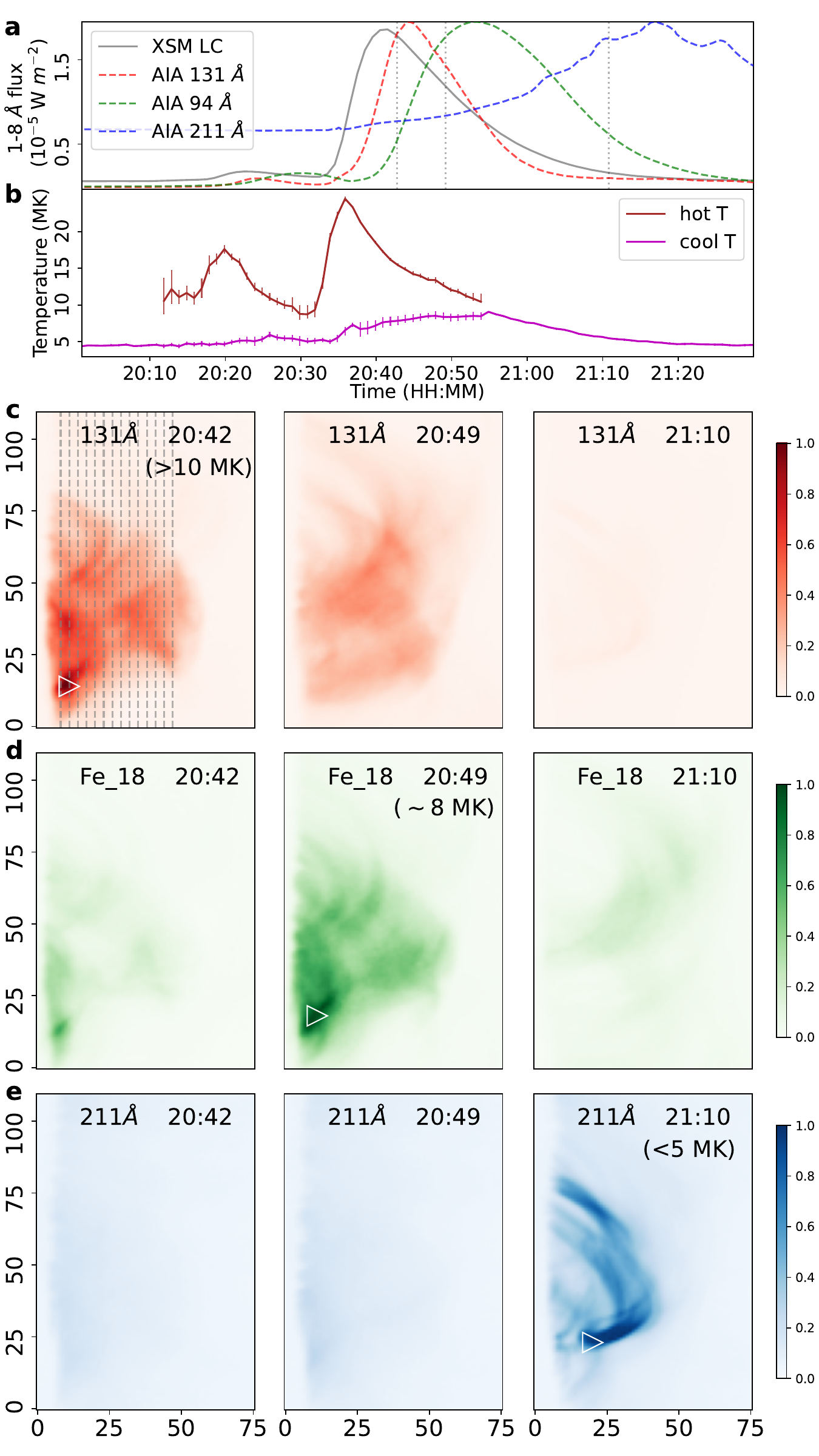}
   \caption{{\bf Observations of the flare: Sol2022-09-17T20:40. $|$}
{\bf a,} X-ray and EUV light curves of the flare observed by XSM in the 1–8 {\AA} range ({\em grey}), and AIA passbands: 131 {\AA} ({\em red}), 94 {\AA} Fe XVIII ({\em green}), and 211 {\AA} ({\em blue}) respectively. The brown and magenta curves with 1-$\sigma$ errorbars in panel b show the measured temperature from the XSM spectral analysis. Two-temperature represents the two components in spectral fitting.
{\bf c–e,} AIA images at three different times indicated by vertical dashed lines in panel {\bf a}. White arrows indicate the location of maximum bightenings and the vertical dotted lines in panel c represent the slits for which AIA light-curves are shown in Figure~\ref{fig_abund_evol}. The X and Y axes are AIA pixel unit.
    \label{fig-flarelc}}
\end{center}
\end{figure}

The XSM 1–8 {\AA} X-ray light curves (Figure~\ref{fig-flarelc}a, gray color) show a small peak ($\sim$20:20 UT) before the main flare peak ($\sim$20:40 UT), caused by a smaller preceding flare.
The red, green, and blue curves represent the light curves in AIA’s 131~{\AA}, 94 {\AA}/ FeXVIII (green), and 211{\AA} (blue) passbands, respectively. 
%The AIA light curves were obtained by integrating emissions from the flaring region, with Fe XVIII contributions extracted from the 94 {\AA} passband using the method of \cite{2013A&A...558A..73D}. 
Peak timings of these light curves vary due to their temperature sensitivities, consistent with prior observations~\citep{Petkaki_2012A&A...547A..25P}.
Figure~\ref{fig-flarelc}c-e provides zoomed views of the flaring loops in different AIA passbands at various times. In the zoomed images, the solar limb is oriented vertically, revealing that both flare footpoints were behind the limb with loops extending above it.

A comparison of AIA light curves from various active regions during the flare, as shown in Appendix Figure~\ref{fig_sup_img-1}, confirms that the flaring emission dominates over other solar activity. This establishes that disk-integrated XSM observations are predominantly influenced by the flare emission, reinforcing its utility in isolating flare dynamics.

\subsection{Time resolved spectroscopy}\label{sec-sectroscopy}

\begin{figure}[!h]
    \centering
    \includegraphics[width=0.5\textwidth]{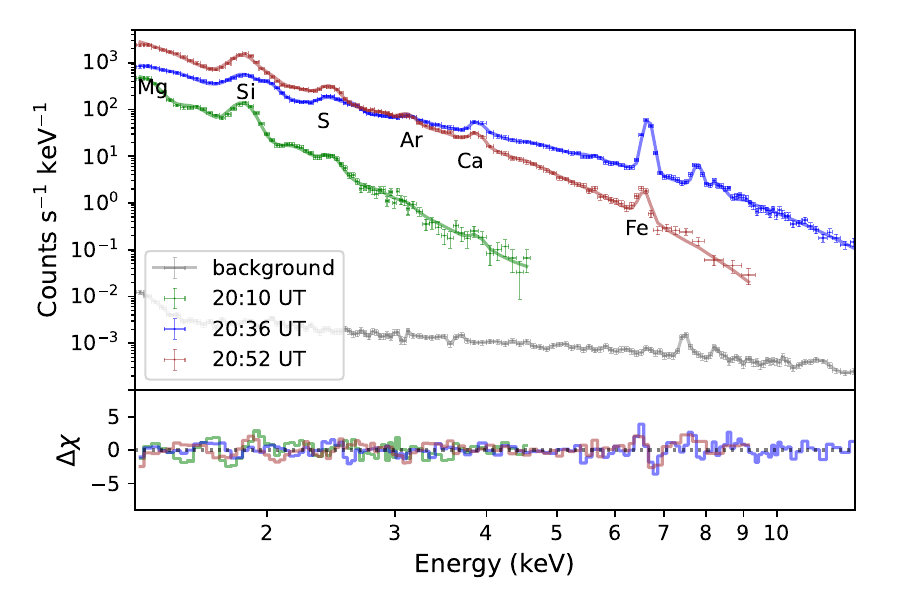}
    \caption{{\bf XSM Spectra} $|$ Green, brown, and blue points with error bars represent the observed XSM spectra at three representative times. Solid lines show the fitted spectra. The gray curve corresponds to the non-solar background spectrum when the Sun was not in the XSM field of view. Lower panel show the residuals.}
    \label{fig_sup_img-2}
\end{figure}

We perform time-resolved spectroscopy using XSM observations to derive the time-evolution of the plasma temperature and elemental abundances. The XSM Data Analysis Software (XSMDAS:~\citealp{Mithun_2021A&C....3400449M}) processes raw XSM data to generate spectra with a one-minute time bin, ensuring sufficient counting statistics for flaring spectra. Spectra below 1.3 keV are excluded due to response uncertainties~\citep{Mithun_2020SoPh..295..139M}, while only the energy range where solar emission dominates over the non-solar background at higher energies is considered for analysis.
The non-solar background spectrum used for this analysis, derived when the Sun was outside XSM's field of view, is available in the XSMDAS calibration database (see Figure~\ref{fig_sup_img-2}).

We utilize PyXspec~\citep{pyxspec_2021ascl.soft01014G}, the Python interface for the standard X-ray spectral fitting package XSPEC\citep{arnaud96}, to analyze the time-resolved spectra. The spectra are fitted using a two-temperature emission model implemented in the XSPEC local model \verb|chisoth|~\citep{Mondal_2021}, which generates model spectra based on tabulated data from the CHIANTI atomic {database v10~\citep{1997A&AS..125..149D, Zanna_2021ApJ...909...38D}.} This model computes spectra for various temperatures, emission measures, and elemental abundances, as detailed in \cite{Mondal_2021}.
{In this process, the abundances are determined relative to the continuum emission, which is dominated by hydrogen at flaring temperatures.}

During the flare, the observed spectra are well-fitted with a two-temperature emission model, whereas in the pre- and post-flare periods, an isothermal emission model adequately explains the spectra. 
Although flaring plasma may exhibit multi-thermal properties, the soft X-ray spectrum is typically approximated using one or two-temperature components. \cite{Mithun_2022ApJ...939..112M} have shown that the inferred elemental abundances remain consistent across various modeling approaches, including multi-thermal Differential Emission Measure (DEM), isothermal, and two-temperature models.

For the spectral fitting, the abundances of all elements are initially set to coronal values from \cite{Feldman_1992PhyS...46..202F}. For each spectrum, the abundances of major elements whose line complexes fall within the energy range being fitted are treated as free parameters, alongside temperature and emission measure. For example, if the spectrum includes Fe line complexes around $\sim$6.5 keV, the abundance of Fe is allowed to vary as a free parameter. 
Abundances of Si, Fe, and S are prominent emission lines during the flare peak. Additionally, Mg line appears partially near the lower energy cutoff of the spectra; however, during the flare peak, blending of this line results in larger uncertainties in its abundance. Ar and Ca line complexes are occasionally visible, but their weak signals result in substantial uncertainties in their abundances when included in spectral fits. Therefore, their abundances are fixed during the fitting process. Treating these abundances as either free or fixed does not affect other fit parameters, consistent with the findings of \cite{Mondal_2021}.

XSM observations, being disk-integrated, include emission from the quiescent non-flaring region. However, the non-flaring emission is significantly weaker than the flare emission (Section~\ref{sec-obs}), making its impact on the inferred flare parameters negligible, even for B-class or C-class flares~\citep{Mondal_2021, Mithun_2022ApJ...939..112M}.

Representative XSM spectra fitted with the model are shown in Figure~\ref{fig_sup_img-2}, while the time-evolution of flare temperature is depicted in Figure~\ref{fig-flarelc}b. During the flare, the model includes two temperature components: the hot component (brown curve) and the cool component (magenta curve). Before and after the flare, the spectra are adequately fitted with a single temperature component.
The evolution of the Fe, Si, and S FIP bias throughout the flare is shown by the blue, brown, and orange curves in Figure~\ref{fig_abund_evol}. To calculate the FIP bias, we take the ratio of the best-fitted abundances to their corresponding photospheric values from \cite{Asplund_2021A&A...653A.141A}. 

%\subsection{Plasma flows}\label{sec-plasmaflows}
\subsection{Plasma dynamics}\label{sec-plasmaflows}

To understand the plasma dynamics, we investigated the image sequences of various AIA passbands.  To measure high-temperature plasma emission, we used the AIA 131~{\AA} passband, which is dominated by the Fe~XXI (128.75~{\AA}, log~$T \sim 7.05$) emission line~\citep{Dwyer_2010A&A...521A..21O}.  
For plasma at around 8~MK, we extracted the Fe~XVIII contribution from the AIA 94~{\AA} passband.  
Cooler plasma ($<$5~MK) was observed in the AIA 211~{\AA} passband, dominated by Fe~XVII (204.7~{\AA}, log~$T \sim 6.6$) and Ca~XVI (208.6~{\AA}, log~$T \sim 6.7$)~\citep{Dwyer_2010A&A...521A..21O}.  
Figures~\ref{fig-flarelc}c--e show images of the flaring loops in these three passbands at different times. 
The arrows indicate representative locations near one leg of the flaring loop, where maximum intensity brightening occurs near the peaks of the light curves in each band, possibly indicating plasma evaporated from the lower atmosphere.
To investigate this further, we examined the peak intensity times at each location along the flaring loops by analyzing the light curves extracted from the slits marked in Figure~\ref{fig-flarelc}c for all three passbands. The leftmost slit is located near the footpoint of the flaring loop, while the rightmost slit is positioned near the looptop. 
The inverted light curves in three AIA passbands for all these slits are shown in Figure~\ref{fig_abund_evol} by red, green, and cyan colors.
The peak times of these light curves indicate that the emission peaks progress from left to right along the slits, which could be due to dense evaporating plasma passing the leftmost slit before reaching the rightmost one. However, without high-resolution spatial data and Doppler measurements, a definitive conclusion cannot be drawn.

\section{Results}\label{sec-results}

\begin{figure*}[ht!]
\begin{center}
    \includegraphics[width=1\textwidth]{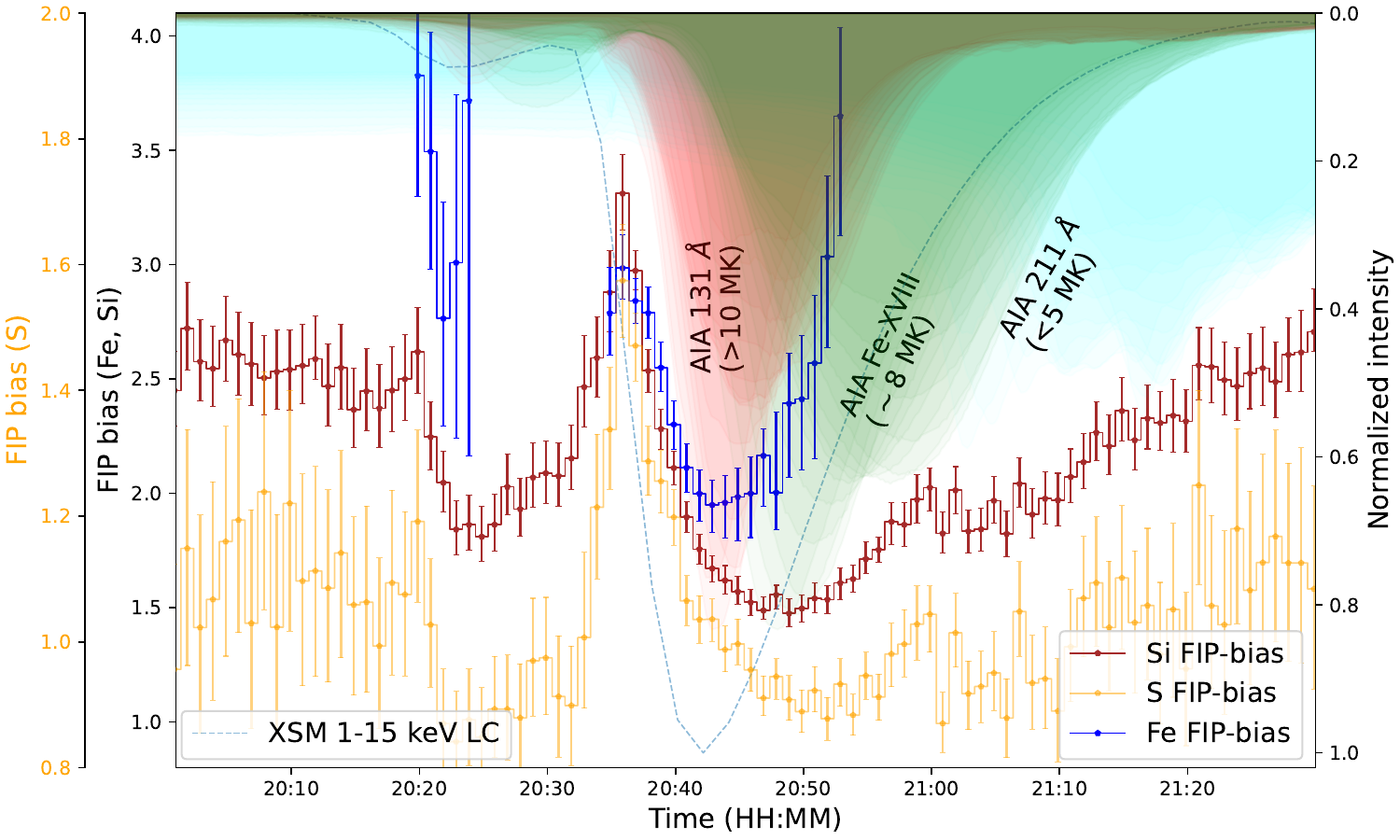}
    \caption{{\bf Correlation between FIP bias variations and plasma flows} $|$ Brown, blue, and orange points with 1-$\sigma$ error-bars represent the variation of the FIP bias for Fe, Si, and S during the flare. Fe FIP bias measurements are available only during periods of intense emission at higher temperatures. The red, green, and cyan filled curves show the normalized flare light curves in the AIA 131 {\AA}, Fe XVIII, and 211 {\AA} passbands, measured along the slits marked in the left panel of Figure~\ref{fig-flarelc}b. Arrows indicate the plasma flows derived from intensity peaks at each slit: right-facing arrows denote upflows, while left-facing arrows represent downflows. The blue dotted line represents the normalized XSM light curve. The y-axis of all light curve are inverted. 
    \label{fig_abund_evol}}
\end{center}
\end{figure*}

Figure~\ref{fig_abund_evol} shows the time evolution of measured FIP bias for Fe (blue), Si (brown), and S (orange). We focus on the primary flare peak (20:35 UT to 21:00 UT), where XSM spectra provide robust statistics for all elements. During the initial rapid heating phase~\citep{Benz_2017LRSP...14....2B} at 20:35 UT, when the temperature peaks (Figure~\ref{fig-flarelc}b), all elements exhibit strong coronal FIP bias, indicating coronal plasma heated by high-energy flaring particles.  This pronounced coronal FIP bias likely correlates with the hot X-ray `onset' interval reported by \citet{Hudson_2021MNRAS.501.1273H} at the beginning of the flare’s impulsive phase, a connection that warrants further detailed investigation. As the flare progresses, the FIP biases decline toward photospheric levels, reflecting plasma evaporation from deeper chromospheric layers, which is consistent with earlier observations~\citep{Mondal_2021,Mithun_2022ApJ...939..112M,Woods_2023ApJ...956...94W}. After the flare peak, the FIP biases recover to coronal values but at differing rates: Fe recovers rapidly just after the X-ray emission peak, while Si lags, maintaining near-photospheric levels briefly before gradually returning to coronal values. The mid-FIP element S shows a similar trend to Si but recovers more slowly during the decay phase. Linear fits to the FIP bias recovery rates yield 0.27 min$^{-1}$ for Fe, 0.06 min$^{-1}$ for Si, and 0.02 min$^{-1}$ for S.  These recovery rates correlate with plasma emission observed in AIA passbands.  Fe’s rapid recovery aligns with $>$10 MK  emission in AIA 131~{\AA} (red light curves in Figure~\ref{fig_abund_evol}), while the slower recovery of Si and S corresponds to $\sim$8 MK Fe XVIII emission (green curves in Figure~\ref{fig_abund_evol}). In the later phase, after 21:00 UT both Si and S recover even more slowly, correlate with  emission observed in cooler $<$5 MK plasma in AIA 211~{\AA} passband.

\section{Discussion}\label{sec-discussion}

In this study, we investigated the relationship between element-specific abundance variations and temperature-sensitive plasma emission during an M-class solar flare. Figure~\ref{fig-flarelc}a shows the flare light curves across different energy passbands, while Figure~\ref{fig-flarelc}a shows the evolution of temperatures derived from XSM spectral analysis. The timing of the emission peaks varies among the passbands due to their distinct temperature sensitivities. XSM’s X-ray flux (gray), which is most sensitive to high-temperature plasma, peaks earlier than the corresponding intensity maxima in AIA’s 131~{\AA} (red), Fe XVIII (green), and 211~{\AA} (blue) passbands. At the emission peak of the AIA passbands (particularly 131~{\AA} and Fe XVIII), the flaring loops exhibit sequential brightening (Figure~\ref{fig-flarelc}c-e), originating near one footpoint and extending into the corona (Section~\ref{sec-plasmaflows}), reaching heights beyond 10 Mm. 
These brightening are first detected between 20:41 to 20:45 UT in the hottest plasma ($>$10 MK) observed in AIA 131~{\AA} (Figure~\ref{fig-flarelc}b), followed by $\sim$8 MK plasma in Fe XVIII (Figure~\ref{fig-flarelc}c) between 20:45 to 20:49UT.

\begin{figure*}[ht]
\begin{center}
    \includegraphics[width=0.99\textwidth]{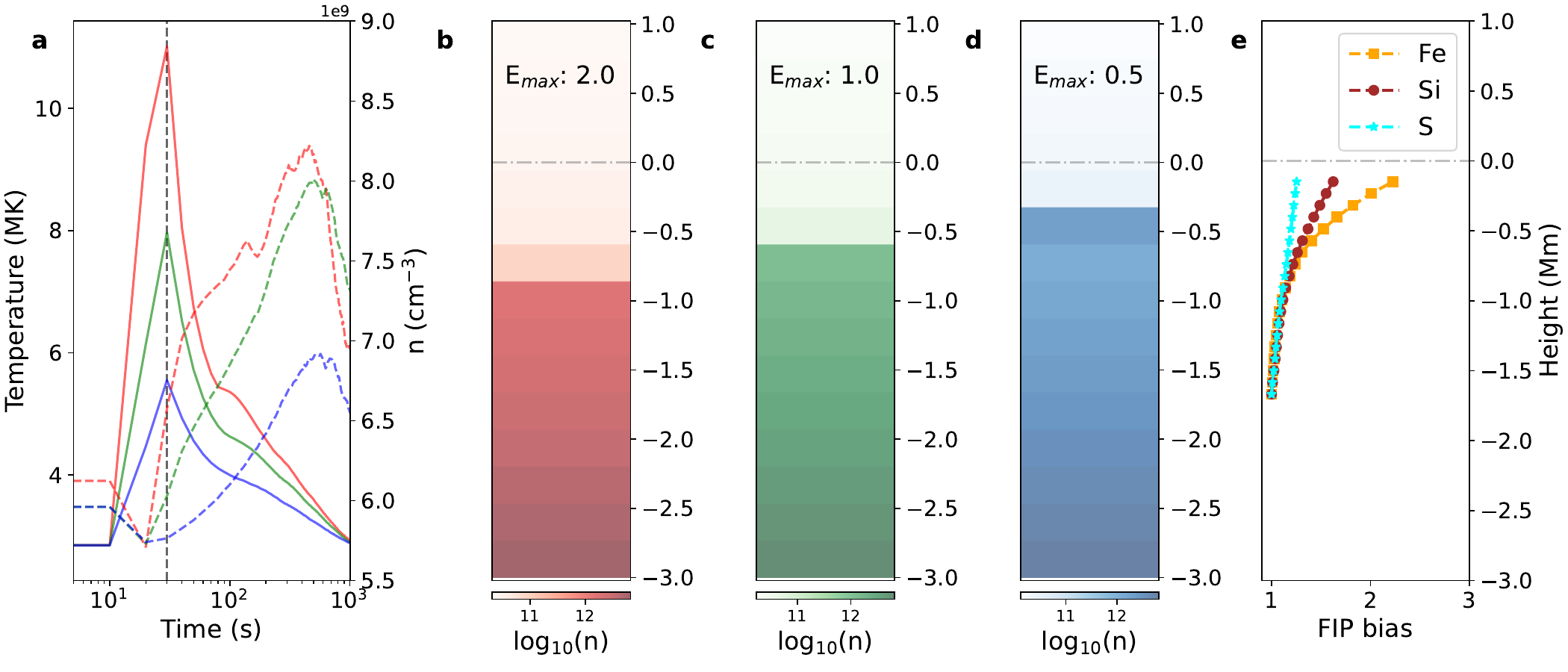}
    \caption{{\bf Simulated coronal loop showing plasma perturbation from different chromospheric heights in response to varying heat deposition.} $|$ {\bf a} Time evolution of coronal temperature (solid lines) and density (dashed lines) for three different simulations with  maximum volumetric heat deposition rates of 2.0 erg cm\(^{-3}\) s\(^{-1}\) (red), 1.0 erg cm\(^{-3}\) s\(^{-1}\) (green), and 0.5 erg cm\(^{-3}\) s\(^{-1}\) (blue) at the top of the chromosphere.  
    {\bf b–d} Density perturbations at different chromospheric depths at the time step indicated by the vertical dashed line in panel {\bf a}, for the three heat deposition cases. The horizontal dashed lines mark the top of the chromosphere.
{\bf d} Variation in FIP fractionation of Fe, Si, and S at different chromospheric depths for a typical closed coronal loop environment due to the ponderomotive force, as estimated by \cite{Laming_2019ApJ...879..124L}.
\label{fig_chevap}}
\end{center}
\end{figure*}

%According to Laming's ponderomotive force theory~\citep{Laming_2004ApJ...614.1063L,Laming_2012ApJ...744..115L} of the FIP effect, the reflection of magnetic waves in the lower atmosphere of a typical closed coronal loop separates ions from neutrals, driving elemental fractionation. 
Figure~\ref{fig_abund_evol} shows the time evolution of the FIP bias of Fe, Si, and S, together with the flaring emissions measured in the AIA passbands. The element-specific FIP bias variations are correlated with the emissions in different passbands (after 20:35 UT), each sensitive to different temperatures (Section~\ref{sec-results}).
The observed correlation between plasma emissions and the FIP bias recovery timescale during the decay phase provides  evidence for the interplay between the flaring process and the FIP effect.
During the impulsive phase (post-20:35 UT), intense heat deposition drives the evaporation of unfractionated plasma from the deep chromosphere, leading to a near-uniform depletion of FIP bias across all elements. As the flare evolves and energy deposition decreases, evaporation arises from progressively higher chromospheric layers, where plasma fractionation becomes increasingly pronounced and varies among different elements according to their ionization balance. 
Figure~\ref{fig_chevap}e shows the FIP fractionation of Fe, Si, and S with chromospheric height, resulting from resonant wave-driven ponderomotive forces in a typical closed coronal loop, as calculated by \citet{Laming_2019ApJ...879..124L}.  
Fe (FIP = 7.87 eV), which exhibits a higher fractionation rate with increasing chromospheric height, is expected to recover its coronal abundance more rapidly -- consistent with the observations shown in Figure~\ref{fig_abund_evol}.
Si (FIP = 8.15 eV), with a slightly lower fractionation rate, recovers slowly, in agreement with observations.
S (FIP = 10.36 eV), which fractionates at a significantly slower rate than Fe and Si, exhibits the slowest recovery.

%
%It is also worth noting that, in the literature~\citep{Laming2015}, sulfur (S) has occasionally been reported to behave like a high-FIP element—exhibiting an inverse FIP bias (i.e., a value less than 1). 
%In Figure~\ref{fig_abund_evol}, S appears to have values slightly below unity; however, given the associated error bars and the fact that the absolute abundance values slightly depend on those of other elements (e.g., O, C, etc) that contribute to the continuum (see Figure 12a of \citealp{Mondal_2021} and the accompanying discussion), it is difficult to definitively conclude whether S is behaving as a high-FIP element in this case.

The interplay between flaring heat deposition and the height of plasma perturbation in the chromosphere is supported by field-aligned hydrodynamic simulations performed with the HYDRAD code \citep{Bradshaw_2003A&A...401..699B,Bradshaw_2013ApJ...770...12B}. We modeled a semi-circular loop of length 60 Mm, with 5 Mm at each end representing the chromosphere. The loop was initialized in hydrodynamic equilibrium with a footpoint temperature of 20,000 K and a density of  10$^{11}$ cm$^{-3}$. At t=10 s, a triangular heating pulse lasting 20 s was deposited at the top of the chromosphere, and the plasma response was sampled every 10 s. 
To explore the effect of heating magnitude, we repeated the simulations for peak heating rates of 2.0 erg cm\(^{-3}\) s\(^{-1}\), 1.0 erg cm\(^{-3}\) s\(^{-1}\), and 0.5 erg cm\(^{-3}\) s\(^{-1}\). Figure~\ref{fig_chevap}a shows the averaged coronal temperature and density evolution, while panels b–d display the snapshot of chromospheric density profiles at  t=30 s for the three heat deposition rates, respectively.

At the highest heat deposition (Figure~\ref{fig_chevap}b), plasma perturbation originates from deeper chromospheric layers, where Fe, Si, and S remain un-fractionated (Figure~\ref{fig_chevap}e). 
For intermediate heat deposition (Figure~\ref{fig_chevap}c), evaporation occurs from progressively higher layers, where Fe (orange curve in Figure~\ref{fig_chevap}e) undergoes greater fractionation compared to Si (brown curve) and S (cyan curve), producing a higher FIP bias for Fe in coronal observations. 
{In contrast, under minimal heat deposition (Figure~\ref{fig_chevap}d), the evaporated plasma originates primarily from the upper chromosphere. 
In this regime, Fe shows the highest degree of fractionation, followed by Si and then S (Figure~\ref{fig_chevap}e) -- trends that align with the observed recovery rates of the FIP bias.
}

%In contrast, minimal heat deposition (Figure~\ref{fig_chevap}d) triggers evaporation from the upper chromosphere, where the plasma reaches lower temperatures, and Fe exhibits the highest fractionation, followed by Si and then S. These trends are consistent with the observed FIP bias recovery rates.

%In conclusion, the variation of abundances during flares, specifically the recovery of the abundances from photospheric to coronal values in the decay phase, could be explained by the flaring mechanism. 

Thus, the recovery of abundances is highly sensitive to the characteristics of energy deposition during flares. Previous studies~\citep{Katsuda_2020ApJ...891..126K, Ng_2024ApJ...972..123N} have reported the inverse-FIP effect during larger flares. Such events may involve a similar mechanism, where flare energy evaporates plasma from much deeper atmospheric layers influenced by the inverse-FIP effect, as described by Laming et al.\citep{Laming_2021ApJ...909...17L}.
Abundance variability also depends on whether it is measured from cooler or hotter plasma emissions. This aligns with findings by \citet{Andy_2024A&A...691A..95T}, who observed reduced abundance variation over time in a flaring loop by analyzing lower-temperature EUV emission line ratios of a high-FIP element (Ar) to a low-FIP element (Ca). 
{Earlier EUV studies (e.g., \citealp{Doschek_2018}) reported coronal abundances at loop top and photospheric at footpoints. In contrast, our study, observing only the coronal portion with both footpoints occulted, shows significant temporal variation in FIP bias and near photospheric abundance during flare peak. This further suggests that X-ray emission from hotter plasma may exhibit a different FIP bias than EUV-emitting plasma. Moreover, while EUV studies often classify S as a high-FIP element in deriving FIP bias, present X-ray observations indicate S behaves more like a low-FIP element, with its abundance varying significantly throughout the flare. Therefore, caution is advised when using sulfur as a reference element for FIP bias determination.}
Furthermore, the model by \cite{Laming_2019ApJ...879..124L} assumes a quasi-static, quiescent chromosphere, which may not be a valid approximation for the highly dynamic conditions of a flaring chromosphere. As a result, the FIP bias distribution in the chromosphere during flares may differ from what is shown in Figure~\ref{fig_chevap}e. Additionally, flare-driven Alfvén waves \citep{Fletcher_2008ApJ...675.1645F} may accelerate the fractionation process, as suggested by \citet{Mondal_2021}, although direct observational evidence for such waves remains lacking.

Looking ahead, the growing modeling efforts (e.g.,~\citealp{Reep_2024ApJ...970L..41R}) together with upcoming observational missions will advance our understanding of the physical origins of abundance fluctuations during solar flares. NASA’s CubeSat Imaging X-ray Solar Spectrometer 
(CubIXSS:~\citealp{Caspi_2023SPD....5420704C}) and the Multi-slit Solar Explorer 
(MUSE:~\citealp{Pontieu_2022ApJ...926...52D}) will provide measurements of elemental abundances and coronal plasma velocities, offering critical insights into the underlying mechanisms. In addition, the Marshall Grazing Incidence X-ray Spectrometer (MaGIXS:~\citealp{Sabrina_2023ApJ}), an slit-less spectrometer sensitive to hot X-ray emission~\citep{Athiray_2019ApJ,Mondal_2024ApJ...967...23M} and capable of measuring abundance maps~\citep{Mondal_2025arXiv250814866M}, is scheduled for a third flight to investigate the physics of abundance recovery during the decay phase of solar flares.

\section{Summary}\label{sec_summary}

We studied the evolution of the FIP bias of Fe, Si, and S during an M-class solar flare. In the impulsive phase, all three elements showed a uniform depletion of abundances from pre-flare coronal values to near-photospheric values, consistent with intense evaporation of chromospheric material in the early phase of the flare. In the decay phase, however, Fe recovered its coronal abundance much more rapidly than Si and S: the Fe recovery timescale correlated with very hot ($>$10 MK) plasma emission, whereas the recovery of Si and S correlated with the emissions of cooler plasma (6–8 MK). These results suggest a link between flare-driven plasma dynamics and the redistribution of elemental abundances. In the decay phase, evaporation originates from progressively higher chromospheric layers, leading to element-dependent recovery rates, as different elements exhibit distinct fractionation behaviors with height depending on their ionization balance.
In conclusion, the recovery of abundances during the decay phase can be explained by the flare-driven evaporation process; however, alternative mechanisms—such as flare-driven Alfvén waves—cannot be ruled out. Future studies that combine detailed modeling with advanced observations will be essential for developing a more complete understanding.

%This framework may also be applicable to magnetically active stars, providing new insights into the interplay between energy input, plasma flows, and elemental fractionation.

%It studied the evolution of FIP bias of Fe, Si, and S during the course of an M class solar flare. In the impulsive phase of the flare, all elemnets show depletion of the abundances from preflare coronal values to near photospheric values, while in the decay phase, Fe recover its coronal abundances in a much quicker rate than Si, and S. The recovery time scale of the Fe coaligne with the emission of very hot ($>$10 MK) plasma emission, while the recovery time scales of Si and S aligned with the cooler (6-8 MK) plasma emissions. From these results we established a possible link between flaring process and plasma dynamics in redistributing elemental abundances during solar flares, shedding light on the underlying mechanisms.  In the impulsive phase, intense evaporation from the deep chromosphere, leading uniform depletion of abundances in all elements, while in the decay phase, evaporation originate from progressively upper chromospheric height, leading to different recovery rate for different elements depending on their fractionation rate in the process of FIP effect. 

%It provides a framework applicable to similar phenomena in magnetically active stars, enhancing our understanding of the interplay between energy input, plasma flows, and elemental fractionation. 

%###################################################################
\vspace{0.5cm}

\section*{Appendix}

\renewcommand\thefigure{\thesection\Alph{figure}}
\setcounter{figure}{0}

\subsection{Flaring Contribution to XSM Observations}

\begin{figure*}[ht]
    \centering
    \includegraphics[width=1.0\textwidth]{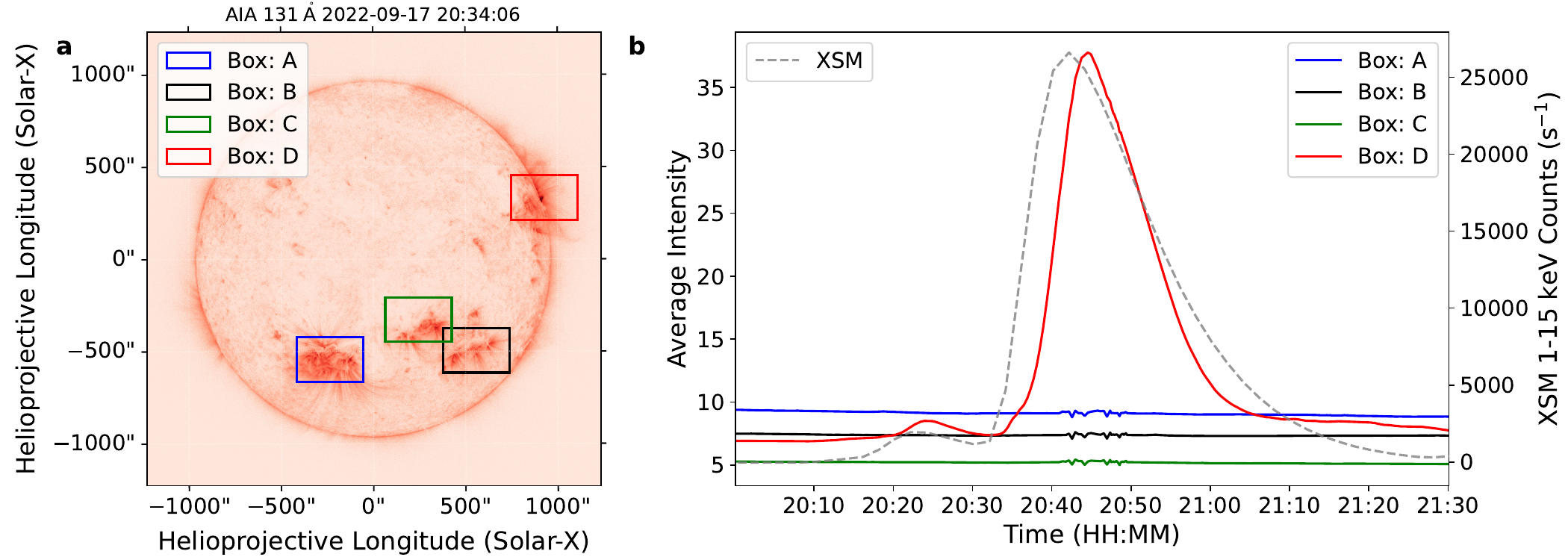}
    \caption{{\bf Location of Flare Sol2022-09-17T20:40} $|$ {\bf a} Full-disk AIA 131 {\AA} image of the Sun, highlighting all major activity. The red box indicates the flare location. {\bf b} AIA 131 {\AA} light curves corresponding to the regions marked in {\bf a}, color-coded to match the boxes. The gray dashed curve represents the XSM X-ray count rate in the 1–15 keV energy range.}
    \label{fig_sup_img-1}
\end{figure*}

Figure~\ref{fig_sup_img-1} shows the contribution of flaring emission in the disk-integrated observation of XSM. The light curves shown in panel (b) indicate that the AIA 131{\AA} emission from the flaring region correlates with the XSM light curve and is much higher compared to the other regions. The time lag between the XSM and the AIA is due to differences in their temperature sensitivities.\\

%\subsection{Animated Plasma Flow Dynamics}\label{sec_appendix_plasmaflow}

%The full dynamic evolution of the flaring plasma and associated flows is presented in Supplementary Movie 1. Panel {\bf a} shows normalized flare light-curves (LC) observed by XSM (gray) and AIA passbands: 131 {\AA} (red), Fe XVIII (green), and 211 {\AA} (blue). AIA light curves correspond to the plasma dynamics across slits marked in panel {\bf b}. Pannel {\bf b--d} depict the evolution of plasma flows in the respective AIA passbands, as labeled. Arrows indicate the flow direction within flaring loops: right-facing arrows represent upflows, while left-facing arrows denote downflows.\label{fig_sup_mov-1}\\

\noindent {\bf Data Availability}\\
The XSM and AIA data used in this study are publicly available at the Indian Space Science Data Center (ISSDC) and the Joint Science Operations Center (JSOC), respectively.
The \verb|chspec| model used for spectral fitting is openly accessible at \url{https://github.com/xastprl/chspec}.
The HYDRAD hydrodynamic simulation code is also publicly available at \url{https://github.com/rice-solar-physics/HYDRAD}.
The data supporting the plots presented in this paper can be obtained from the corresponding author upon reasonable request.

%and its python interface Pydrad are publicly available at ..

%\noindent {\bf Acknowledgements}\\

\acknowledgments{
We thank Dr. J. Martin Laming for providing the data used in Figure 4e and for his insightful comments on the role of resonant frequency in FIP fractionation.
BM's research was supported by an appointment to the NASA Postdoctoral Program at the NASA Marshall Space Flight Center, administered by Oak Ridge Associated Universities under contract with NASA.
BM tanks to Dr. Anshika Banshal, for her thoughtful suggestions that helped improve the manuscript. 
This work makes use of data from SDO/AIA and XSM/Chandrayaan-2. The XSM instrument was developed by the engineering team at the Physical Research Laboratory (PRL), led by Prof. S. Vadawale and Dr. M. Shanmugam, with support from several ISRO centers. We also acknowledge the use of the CHIANTI atomic database, a collaborative project involving George Mason University, the University of Michigan, the University of Cambridge, and NASA Goddard Space Flight Center.}

{\textit{Facilities:} Chandrayaan-2/XSM SDO/AIA.}

{\textit{Software:} Astropy~\citep{Astropy2018AJ....156..123A}, IPython~\citep{ipython_2007CSE.....9c..21P},matplotlib~\citep{matplotlib_2007CSE.....9...90H}, NumPy~\citep{2020NumPy-Array}, scipy~\citep{2020SciPy-NMeth}, SunPy~\citep{sunpy_community2020}, SolarSoftware~\citep{Freeland_1998}, 
Chianti~\citep{Dere_1997A&AS..chianti,chiantiV10_Zanna2020}}

\bibliographystyle{aasjournal}
\bibliography{references_xsm}

\end{document}